\pdfoutput=1
\documentclass[conference]{IEEEtran}
\renewcommand\IEEEkeywordsname{Keywords}
\usepackage[T1]{fontenc}
\usepackage[scaled]{helvet}
\usepackage{luximono}

\usepackage{cite}

\ifCLASSINFOpdf
   \usepackage[pdftex]{graphicx}
   \graphicspath{{../pdf/}{../jpeg/}{./Figures/}}
   \DeclareGraphicsExtensions{.pdf,.jpeg,.png}
\else
\fi
\usepackage{nth}
\usepackage{amsmath}
\usepackage{breqn}
\usepackage{caption}
\usepackage{subcaption}
\usepackage{epstopdf}
\usepackage{enumerate}

\usepackage{algorithmic}
\usepackage{algorithm}

\usepackage{array}
\usepackage{url}

\usepackage{nth}
\begin{document}
\newcommand\pN{\mathcal{N}}	
%
\title{Genetic Algorithm-based Mapper to Support Multiple Concurrent Users on Wireless Testbeds}


\author{
\IEEEauthorblockN{Yaser A. Elnakieb}
\IEEEauthorblockA{Electrical and Computer Eng. Dept.\\ 
	Virginia Tech MENA\\
Email: ynakieb@vt.edu}
\and
\IEEEauthorblockN{Michael Azmy}
\IEEEauthorblockA{Computer and Systems Eng. Dept.\\
	Faculty of Engineering,\\
	Alexandria University, Egypt\\
Email: michael.azmy@alexu.edu.eg}
\and
\IEEEauthorblockN{Mustafa ElNainay}
\IEEEauthorblockA{Computer and Systems Eng. Dept.\\
 Faculty of Engineering,\\
  Alexandria University, Egypt\\
Email: ymustafa@alexu.edu.eg}
}

\maketitle
\begin{abstract}
	Communication and networking research introduces new protocols 
	and standards with an increasing number of researchers relying on real experiments rather than simulations to evaluate the performance of their new protocols. A number of testbeds are currently available for this purpose and a growing number of users are requesting access to 
	those testbeds. This motivates the need for better utilization of the testbeds by
	allowing concurrent experimentations. In this work, we introduce 
	a novel mapping algorithm that aims to maximize wireless testbed 
	utilization using frequency slicing of the spectrum resources. The 
	mapper employs genetic algorithm to find the best combination 
	of requests that can be served concurrently, after getting all 
	possible mappings of each request via an induced sub-graph 
	isomorphism stage. The proposed mapper is tested on grid testbeds and randomly generated topologies. The solution of our mapper is compared to the optimal one, obtained through a brute-force search, and was able to serve the same number of requests in 82.96\% of testing scenarios. Furthermore, we show the effect of the careful design of testbed topology on enhancing the testbed utilization by applying our mapper on a carefully positioned 8-nodes testbed. In addition, our proposed approach for testbed slicing and requests mapping has shown an improved performance in terms of total served requests, about five folds, compared to the simple allocation policy with no slicing. \\
\end{abstract}
\vspace{-0.5mm}
\vspace{-0.5mm}
\IEEEpeerreviewmaketitle

\vspace{-0.5mm}
\section{Introduction}
\vspace{-1mm}
Developing new communication and network protocols and standards is the focus of tremendous number of research studies. To evaluate the performance of theses protocols and standards, researchers have been conducting theoretical analysis, simulation, and/or real experimentation on testbeds. Theoretical analysis and simulations suffer in most cases from over simplifications. Moreover, simulations are affected by the complexity of the 
simulation framework. Therefore, it is more credible for researchers to verify 
their ideas on real testbeds. Many academic testbeds are currently available for this purpose, such as Orbit \cite{raychaudhuri2005overview}, GENI \cite{berman2014geni}, Emulab \cite{Emulab}. An increasing number of users are demanding access to those testbeds which necessitates the sharing
of testbed resources among multiple concurrent users \cite{stavropoulos2015design}. While there are many techniques on wired testbeds for 
resources slicing to share/isolate different experiments efficiently, the problem 
is more challenging in wireless testbeds. Therefore, there is a need for efficient automated mapping and scheduling algorithms for wireless testbeds that aim to maximize the number of concurrent non-conflicting slices.\\
Currently, most of the mapping techniques introduced in the 
literature target wired testbeds. The 
problem of assigning nodes to user requests in a wired-connected testbed 
without violating link constraints is known to be NP-hard \cite{andersen2002theoretical}. 
Fair sharing of overall testbed resources (bandwidth, memory, 
and computational capabilities) among users is a big 
challenge as well. The main difference between the mapping problem in 
a wired versus a wireless environment is in the slicing of the 
testbed. While wired resources can be sliced efficiently, and 
interfaces can be virtualized with minimal effect, sharing a 
wireless interface may affect the sharing of interfaces on other 
nodes because of the wireless resources' inter-dependencies.\\
To guarantee the isolation of experiments, wireless 
testbeds such as Orbit allocates the entire testbed to one user per time (until now). Other wireless testbeds that enables sharing as 
NITOS \cite{NITOS}, relies on on-demand static allocation based on 
frequency slicing. This paper presents a novel mapping technique based on genetic algorithm that considers the testbed topology and utilizes spectrum slicing and resource virtualization to map multiple concurrent users maximizing the wireless testbed utilization. 
The contributions of this paper are: introducing an automated 
mapping algorithm that benefits from virtualization and 
spectrum slicing to maximize the number of concurrent users 
on a wireless testbed, a quantitative evaluation to highlight the advantages and performance of the algorithm, and revealing the possible extensions of our mapper.\\
The rest of this paper is organized as follows. A review of related work is discussed in Section \ref{sec:related_work}. The testbed network model and the tackled problem are described in Section \ref{sec:network_model}. The proposed genetic algorithm-based mapper is detailed in Section \ref{sec:algorithms}. Evaluation results of the algorithm are shown on Section \ref{sec:evaluation}. Finally, conclusion and future work are presented in Section\ref{sec:conc}.
\vspace{-3.5mm}

\section{Related Work}
\label{sec:related_work}
\vspace{-1mm}
The wireless testbed mapping problem is similar to the previous works on Virtual Network Embedding (VNE) on shared wired testbeds \cite{ricci2003solver , chowdhury2009virtual , lischka2009virtual}. However, on a wired topology, requests generally consists of bandwidth requirements between node pairs and node capacity, without accounting for interference on a wireless medium or its effect on the maximum number of allowed virtual machines on a node. Thus, those algorithms mainly map a required virtual topology as source/destination pairs. On the other hand, mapping multiple concurrent users on a wireless testbed differs in how they can share the wireless medium itself. \\
The standard VNE problem is known to be NP-hard because of its constraints \cite{andersen2002theoretical}. Many approaches were introduced in the literature to solve this problem. Yu et al. \cite{yu2008rethinking} proposed a two-stage solution in which they first map virtual nodes then map virtual links, assuming that path splitting is supported by the underlying network. They then employ path migration to periodically re-optimize the utilization of the substrate network. Chowdhury et al. \cite{chowdhury2009virtual} considered location requirements as well and solved the problem using Mixed Integer Program (MIP) formulation, offering better coordination between the two stages. \\
Other approaches as vnmFlib network mapping library \cite{lischka2009virtual}, implemented by Lischka and Karl, used the fact that the network mapping problem resembles the subgraph isomorphism detection, and developed a backtracking algorithm based on the VF2 algorithm \cite{cordella2004sub} used in the pattern recognition community for finding subgraph isomorphisms in large graphs. We inherit this approach to pre-find the possible mappings for each requested topology as a first stage of our mapper algorithm. VnmFlib maps virtual nodes and links in a single stage with better efficiency. VF2x \cite{yin2012vf2x} was proposed to enhance vnmFlib and solve some of its limitations. \\
The Emulab testbed mapping algorithm, Assign \cite{ricci2003solver}, considers the online embedding problem, where mapping decisions are taken for each incoming request, with bandwidth constraint alongside constraints on exclusive use of nodes. It categorizes resources into equivalence classes to reduce search space and uses simulated annealing to find a good match. Singhal thesis work \cite{singhal2008evaluation} targets network virtualization to allow more than one user to share the same wireless node, using a User Mode Linux (UML) network virtualization technique, which he described and evaluated.\\
In \cite{mahindra2008space}, Mahindra et al. provide a comparison between two types of testbed slicing: the usage of flexible time division versus space separation for small sets of nodes, favoring space over time division. The spectrum slicing is introduced in \cite{anadiotis2009new}and \cite{anadiotis2011towards} in the context of wireless standard 802.11, intending to provide each experiment a "spectrum slice"' that minimally interferes with other slices, through the use of orthogonal channels. The authors use a spectrum slicing technique on Wi-Fi nodes that can be used on OMF-based testbed \cite{rakotoarivelo2010omf}, targeting Orbit. This spectrum sharing technique is employed for slicing testbed resources in our CRC testbed developed in Alexandria University, Egypt\cite{CRCtestbed}.\\
Although many works target virtualization of the wireless interfaces \cite{ bhanage2010splitap, hong2012picasso, wang2011fully, nicholson2010juggler, rivera2010virtualization, liang2015wireless}, most algorithms add limitations to the type of experiments that can be performed, such as delay, overhead, time separation, space separation, transmission power, synchronization, limited topologies, and many others. Due to the difficulties of implementing general solution for interface virtualization, we rely on assigning each of the wireless interfaces of a node to at most one user as a basic assumption in our testbed nodes modeling. Hence, a node, not interface, virtualization is allowed. In addition, the testbed spectrum resources are shared between users by assigning different wireless channels to distinct users whenever possible. This virtualization scheme maximizes testbed utilization while achieving separation between various experiments with no added limitations. Our work is the first to our knowledge that maps users' requests to enable concurrent experiments on wireless testbeds, assigning their requested topology, interfaces, channels and other preferences with no human intervention. It can be applied on existing grid or non-grid testbeds.

\section{Testbed Model And Problem Description}
\label{sec:network_model}
\vspace{-0.5mm}
The target of our mapper is to maximize the testbed utilization by determining the best subset of requests that can coexist concurrently without interfering with each other. The problem is to find which subset of nodes can serve a user's requested topology, and at the same time, fit with as many other users as possible using the spectrum and hardware resources of the testbed. Furthermore, a priority for each request is included in our utility function that may favor serving one request among others (for example, one belonging to senior researchers group versus students).\\
The mapper is implemented as two consecutive stages, the first stage is the induced subgraph isomorphism stage which finds all the possible mappings of every requested topology to our testbed. The second stage is a heuristic search using genetic algorithm to find the combination of requests and mappings that lead to the best utilization. Our algorithm is described in more details in the next section.
\vspace{-0.5mm}
\subsection{Testbed physical model}
\vspace{-1mm}
For our mapping problem, the testbed is viewed as group of $N$ wireless nodes, each having $i$ different types of wireless interfaces (for example, if a node has one WiFi card and two types of software defined radio interfaces, then $i$ = 3). A testbed reservation request specifies the number of nodes, their specifications, and connectivity. Each node can be either a physical node, with all its resources and network interfaces, or a virtual node with slice of computational and memory resources and one or more of the node's wireless interfaces. The developed reservation system benefits from virtualization to allow multiple concurrent testbed users if possible. A node can be shared between up to $i$ virtual machines depending on the requests details, and which interfaces are requested.\\
The topology of the testbed is represented as a square $N \times N$ binary connectivity matrix, with each element $(j,k)$ representing the link between nodes $j$,$k$ , i.e., whether nodes $j$ and $k$ are within the wireless range of each or not. This can be determined empirically through a series of transmission and reception between the testbed nodes to identify the coverage of each node, simplified by defining a threshold for power received for a connected link. Another matrix is used to identify the interfaces that satisfy this connectivity matrix. These two matrices is used to determine which nodes/interfaces can serve a particular request with specific topology and frequencies.\\
Each of the $i$ interfaces has a number of orthogonal channels depending on its type, defined by $max\_ch_i$ , and the channels requested by all requests served in a time-slot cannot exceed that limit to avoid interference. We allow frequency slicing, i.e. sharing the spectrum between different requests when interference can be avoided.
\vspace{-0.5mm}
\subsection{Requests}
\vspace{-1mm}
We have a set $R$ of $n$ requests demanding testbed resources in a time-slot, where $n = |R|$. A set of parameters are associated with each request $r$ to identify its details. These parameters are: requested time-slots $TS_r$, priority of the request $P_r$, requested wireless interfaces $Int_r$, requested frequency channels per interface $Ch_r$, the number of requested nodes $N_r$, whether they are virtual or physical $Type_r$, and the required topology represented by another connectivity matrix $G_r$. A user can request either specific fixed channels, or a number of flexible channels, for each requested interface. Table \ref{request_table} summarizes those notations for use in next sections.
\begin{table}[!t]
	\centering
	\caption {Request's Mathematical Notations}
	\vspace{-1mm}
	\label{request_table}
	\begin{tabular}{|p{1cm} |p{6.5cm}|}
		\hline
		Symbol                          & Description    \\ \hline

		$N_r$ & The number of requested nodes \\ \hline		
		$Type_r$ & Whether the nodes are physical or virtual  \\ \hline
		$Int_r$   & Requested wireless interfaces \\ \hline
		$Ch_r$  & Frequency channels requested per interface \\ \hline
		$G_r$ & The required topology represented by $N_r \times N_r$ connectivity (adjacency) matrix \\ \hline
		$TS_r$  & Number of requested time-slot(s)        \\ \hline
		$P_r$   & Priority of the request as assigned by the system depending on the user category/priority.\\ \hline
	\end{tabular}
	\vspace{-4mm}
\end{table}
\vspace{-0.7mm}
\section{GA-based Mapper}
\label{sec:algorithms}
In this section, the two stages of our proposed mapping algorithm are described.
\vspace{-3mm}
\subsection{Generating Candidate Mappings}
\vspace{-0.5mm}
The first stage of the mapping algorithm is to generate candidate mappings for each requested topology using induced subgraph isomorphism techniques. The induced subgraph isomorphism problem can be defined as follows: given $H$ and $G$ graphs, determine whether there is an isomorphic mapping from $H$ to $G$, and find such mappings if exist. The difference of induced from general non-induced subgraph isomorphism problem is that the absence of an edge (link) in $H$ implies that the analogous edge in $G$ is also absent. In the non-induced problem those edges in $G$ can be present. The induced subgraph isomorphism fits our needs in wireless testbeds as we want to provide the requested links of the user topology and avoid potential interference of any other links.\\ 
We use igraph \cite{igraph} library to get all isomorphic induced sub-graphs of each request. The subgraph $H$ represents the requested topology($G_r$), while $G$ denotes the testbed connectivity. After this step, several mappings are available - if exist - associated with each $G_r$. Those distinct mappings are used to maximize the odds of coexistence between different requests.
\vspace{-4mm}
\subsection{Genetic Algorithm}
\vspace{-1mm}
The second stage of the mapper is the genetic algorithm-based stage (GA-based stage). After getting all the candidate mappings of each request, we apply genetic algorithm for choosing the subset of requests that can run together without conflicts while satisfying the constraints. The details of the GA-based stage of the mapper are described in the following subsections.
\subsubsection{Chromosome Structure}
A GA chromosome consists of $n$ genes, each representing a particular request, where $n$ is the total number of requests.
Each gene consists of two fields:
\begin{itemize}
	\item A bit representing whether the corresponding request shall be allocated or not in the current solution.
	\item An integer representing a pointer to a mapping chosen from the candidate mappings of this request as determined from the first stage.
\end{itemize}
For example, in Table \ref{chromosome_table}, $r_1$ and $r_n$ will both be served with their \nth{3} and \nth{6} mapping respectively, while $r_2$ with its \nth{4} mapping will not be served.
\begin{table}[!t]
	\renewcommand{\arraystretch}{1.2}
	\caption{Chromosome Structure}
	\vspace{-1mm}
	\label{chromosome_table}
	\centering
	\begin{tabular}{ c | c | c | c }
		\hline
		$r_1$ & $r_2$ & ... & $r_n$ \\ \hline
		1 	  & 0 	  & ...	& 1 	\\ \hline
		3 	  & 4	  & ...	& 6 	\\
		\hline
	\end{tabular}
	\vspace{-4mm}
\end{table}
\subsubsection{Initialization}

First, we generate the population of chromosomes with random values. Each gene (request) in the chromosome is initialized randomly for both its existence and mapping. The same gene on different chromosomes can be linked with different mapping of its candidate mappings (if there are many).\\
Then, the requested frequencies by all requests are checked for any conflicts. If no conflict is indicated for all requests, then this step is not needed to be done for any subgroup of the requests later during the algorithm. If a conflict is indicated, then only subset of the requests can be served together and this check has to be performed for each chromosome (subset of requests) in every iteration of the GA.
\subsubsection{Selection}
The fitness function in eq.~\ref{fitness} reflects the priority of the request, the requested time duration, and the penalties resulting from any conflicts. In our case, the genetic algorithm tries to find the chromosome which minimizes that fitness function.
Two types of penalties are calculated: resource penalties and channel penalties. If multiple requests ask for the same resource (node or interface on a particular node), the number of resource conflicts is increased. Moreover, if multiple requests demand the same frequency channel, the number of channel conflicts is increased. The penalties term is the multiplication of these conflicts by a large number.

The fitness function is,
\vspace{-1mm}
\begin{equation}
\label{fitness}
Fitness =  Penalties + \sum_{r=1}^{n} (\frac{w_1}{P_r} + w_2 \times TS_r)
\end{equation}
\vspace{-5mm}
\begin{dmath}
Penalties = \sum_{conflicts} (Resource\_conflicts + Channel\_conflicts) \times Large\_number
\end{dmath}
where $w_1$ and $w_2$ are adjustable parameters used to indicate the relative importance of a term over another. In order to understand this formulation, let us consider the case when all requests have equal priorities, have the same time preferences and no penalties occur. In this case, the problem turns to optimization problem that aims to maximizing the number of requests being served on this slot. The fitness function also allow us to serve a request which consumes larger resources or longer duration if it has higher priority than other request(s).
\subsubsection{Genetic Operations}
The two primary genetic operations, mutation and crossover, are applied to chromosomes randomly with probabilities $p_m$ and $p_c$ respectively.
\begin{itemize}
	\vspace{-0.5mm}
	\item Mutation: Two types of mutations are applied, each with probability 0.5:
	\begin{itemize}
		\vspace{-0.5mm}
		\item Mutation 1: randomly toggle the first field boolean of a gene from 0 to 1 and vice versa keeping the same mapping.
		\item Mutation 2: randomly change the mapping of a request to another one of its candidate mappings.
	\end{itemize}
	\vspace{-0.5mm}
	\item Crossover: Apply a single-point crossover between chromosome pairs.
\end{itemize}

\subsubsection{Termination}
When a maximum number of generations is reached, or no significant relative change in the fitness over a predefined stall generations happens, the algorithm is terminated and the best solution that corresponds to the best chromosome of the last generation is determined.

\vspace{-0.5mm}
\section{Evaluation}
\label{sec:evaluation}
\vspace{-1mm}
In the current section, the simulation framework and the different scenarios used to evaluate the performance of the proposed mapper are described and the results are presented. 

\vspace{-1mm}
\subsection{Simulation Framework}
\vspace{-1mm}
In order to evaluate the performance of our GA-based mapper, we setup a simulation environment for different testbed sizes and parameters. We compare our mapping algorithm to the optimal solution found using the exhaustive search. For this purpose, a brute-force search algorithm (BF) was implemented to find the best possible solution, optimum, given a particular subset of requests.  The algorithms are evaluated and compared under multiple testbed sizes and mappings size limits. Moreover, the revenue of applying our GA-based mapper with enabled spectrum slicing and resources virtualization over a simple allocation policy with no slicing is demonstrated while using all possible mappings provided by the induced graph isomorphism algorithm.\\
The performance metrics considered are:
\begin{enumerate}[i)]
	\item \textit{Number of served requests,} the average number of accepted requests by the mapper to be served.
	\item \textit{Optimality,} the percentage of solutions obtained by our mapper that were identical to those found through brute-force search for the optimal solutions. 
	\item \textit{Slicing revenue,} the increased testbed utilization, in terms of the number of concurrent users that our mapper can allocate relative to the simple allocation policy with no slicing (one user).  
\end{enumerate}

\vspace{-2mm}	
\subsection{Experimental Set-up}
\vspace{-1mm}
\begin{table}[!t]
	\vspace{1.5 mm}
	\centering
	\caption {Values of the parameters used for evaluation}
	\vspace{-1mm}
	\label{parameters_table}
	\begin{tabular}{|p{1.7cm} |p{6cm}|}
		\hline
		Parameter                          & Distribution/Value    \\ [0.5ex] \hline
		
		$pop-size$ & 60 \\ [0.5ex] \hline        
	       
	  $p_m$ & 0.2 \\ [0.5ex] \hline        
		$p_c$ & 0.8 \\ [0.5ex] \hline
			$max-iter$ & 500 \\ [0.5ex] \hline 
			$i$ & 3 interfaces \\ [0.5ex] \hline        
		$Max\_ch_i$ & 13 for $i=1$, 13 for $i=2$, and 40 for $i=3$. \\ [0.5ex] \hline        
		$Ch_r$  &  $\sim \pN(\mu, \sigma^2)$ where $\mu = 0.25 \times Max\_ch_i$, and $\sigma = 1/6 \times Max\_ch_i$. The distribution is limited to half of $Max\_ch_i$. \\ [0.5ex] \hline        
		$P_r$ & $\sim \textit{U}(1, 5)$.\\ [0.5ex] \hline        
		$TS_r$  & $\sim \pN(\mu, \sigma^2)$ where $\mu = 2$, and $\sigma = 5$.\\ [0.5ex] \hline
		$Int_r$  & $\sim \textit{U}(1, 3)$ \\ [0.5ex] \hline	
	\end{tabular}
	\vspace{-5mm}
\end{table}
To evaluate our mapping algorithm, random input data are generated to represent several users with variable request preferences. This subsection describes the experimental setup and the statistics of the parameters used.\\
A wireless testbed physical topology is expressed as graph with nodes represent the testbed nodes and links represent the connectivity of nodes. This information is embedded as a connectivity/adjacency matrix. For the experiments, testbed nodes were assumed to assemble either a grid topology or an arbitrary randomly generated topology. In grid testbeds, each node can only see one neighbor in each direction, and in random topology, the number of connections are random. Different testbed sizes have been used of 3$\times$3 grid (9 nodes), 4$\times$4 grid (16 nodes), 5$\times$5 grid (25 nodes), and 6$\times$6 grid (36 nodes).\\
The number of nodes of each request $N_r$ were picked from a uniform distribution with mean roughly equals to half of the available nodes on smallest testbed (between three and five nodes). For each request, we generate a random adjacency matrix $G_r$ representing its requested topology. This range of number of nodes per request is selected as when the number of nodes exceeds five, most of the randomly generated topologies does not fit on the testbed as indicated from the sub-graph isomorphism stage.\\
The requests parameters $Ch_r$, $P_r$, and $TS_r$ where randomly generated for each request to exhibit different users' categories. The number of requested channels $Ch_r$ were selected from a normal distribution limited to half of the maximum number of channels available at most. Five different categories of users were assumed, and hence five uniformly distributed priorities. Moreover, each request demands a random number of time-slots distributed normally around two slots with a high standard deviation of five. The population size of the GA is 60 and the crossover rate $p_c$ and mutations rate $p_m$ were set to 0.8 and 0.2 respectively. This relatively high mutation rate was used to avoid being trapped in local optima. The GA is terminated after 500 iterations (no further improvement has been found for using more iterations). Table \ref{parameters_table} summarizes the parameters values and distributions used for evaluation.\\
For the first scenario, comparison between our mapper solution and the optimal obtained by brute force search, the number of mappings of a topology $G_r$ produced by the first stage of the mapper will be limited. We tried limiting the solutions to ten and fifteen mappings only, for each incoming request. In addition, the number of users demanding testbed resources $|R|$ is fixed to five. These were done to shrink the search space as increasing it would require enormous computation resources and large amount of time for a single run of the brute-force implementation.
Two settings were tested: first, with no virtualization enabled, i.e. $Type = 0$. Second, enabling requests demanding virtual nodes, $Type = 1$. Each experiment is repeated $10$ times, with different random inputs, and the average value and variance of these repetitions are presented in the simulation results of the following figures.\\
Then, we do the simulations for the second scenario, considering all possible mappings for each incoming request and running the GA-based mapper. The effect of varying testbed sizes will be shown, as well as varying the number of requests on the slicing revenue. Similarly, each experiment is repeated $10$ times, and the average value of these repetitions is presented in the simulation results.
Furthermore, the execution time of both GA and BF algorithms are depicted.
\begin{figure*}[t]
	\centering	
	\begin{subfigure}[t]{0.45\textwidth}
		\centering
		\includegraphics[width=\textwidth]{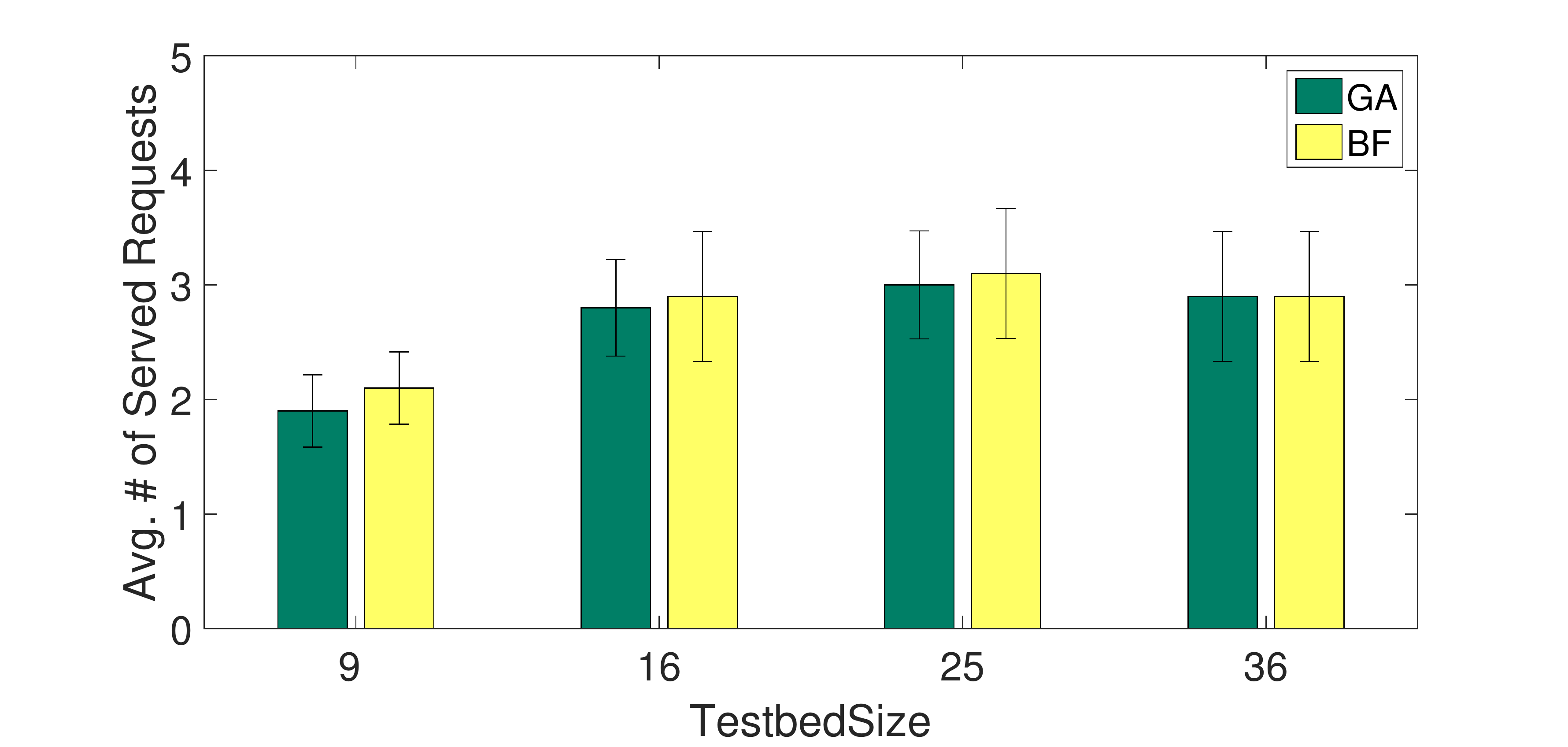}
		\caption{Physical requests, 10 Mappings.}
		\label{fig2:a}
	\end{subfigure}
	\begin{subfigure}[t]{0.45\textwidth}
		\centering
		\includegraphics[width=\textwidth]{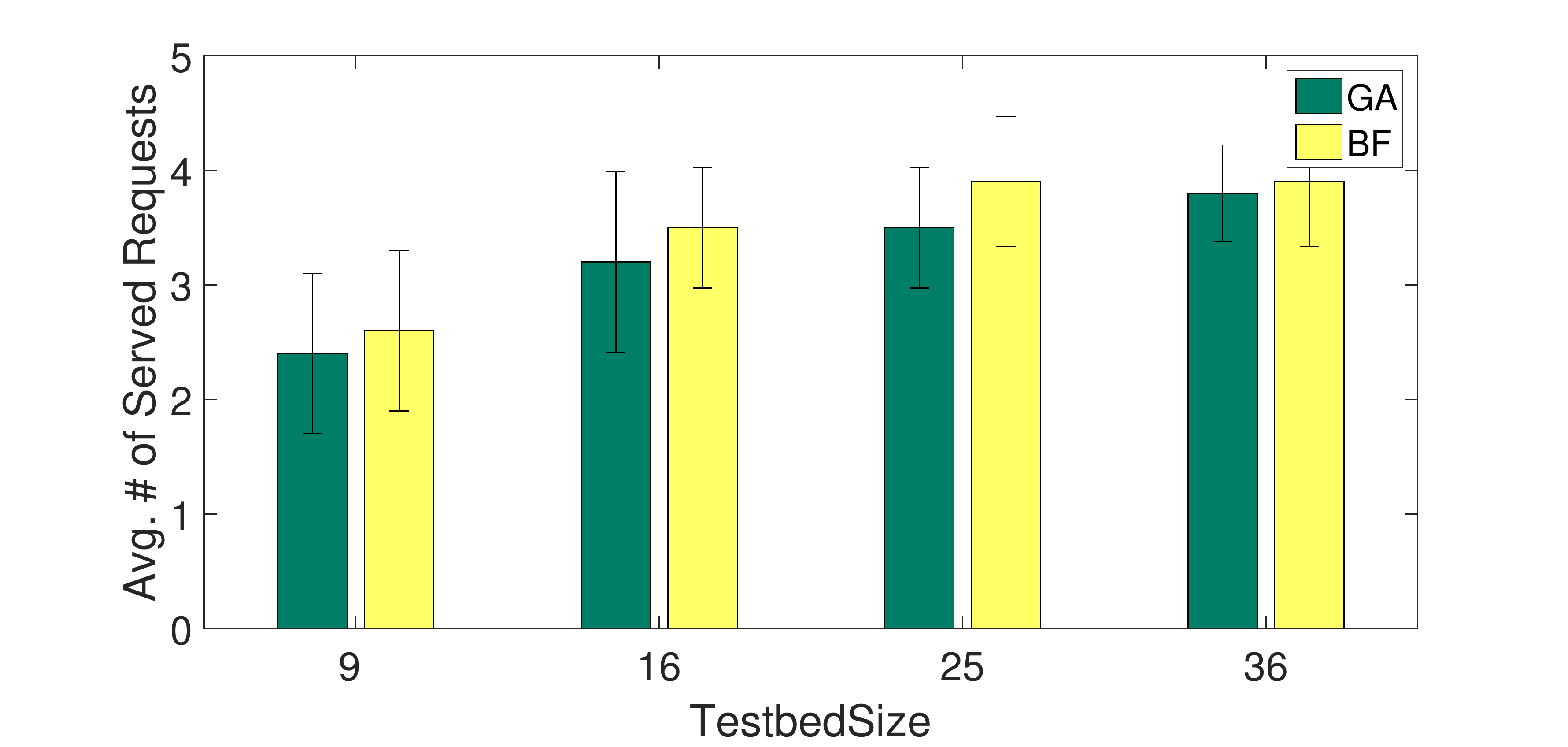}
		\caption{Virtual requests enabled, 10 Mappings.}
	\end{subfigure}	
	\begin{subfigure}[t]{0.45\textwidth}
		\centering
		\includegraphics[width=\textwidth]{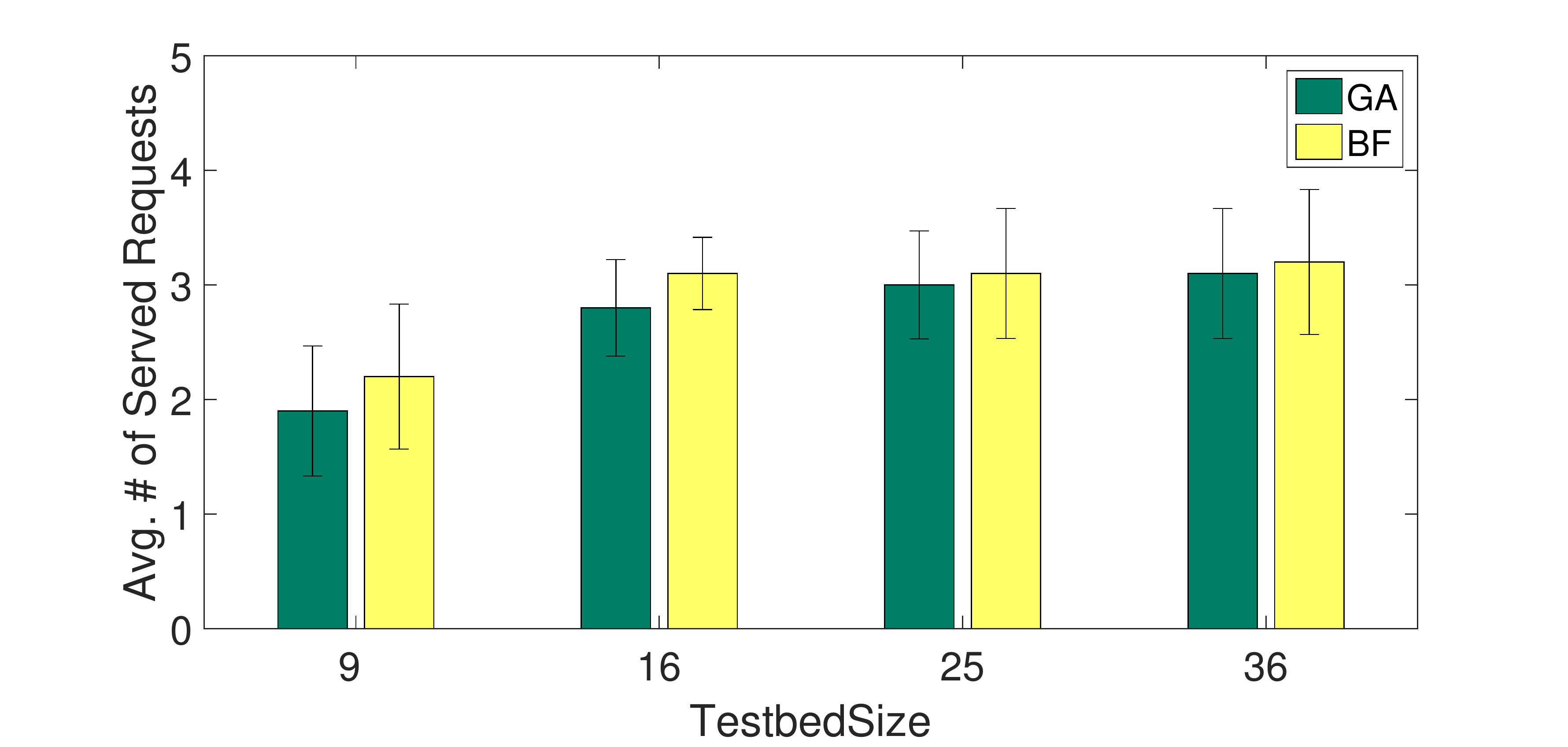}
		\caption{Physical requests, 15 Mappings.}
		\label{fig2:c}
		\vspace{-1mm}
	\end{subfigure}
	\begin{subfigure}[t]{0.45\textwidth}
		\centering
		\includegraphics[width=\textwidth]{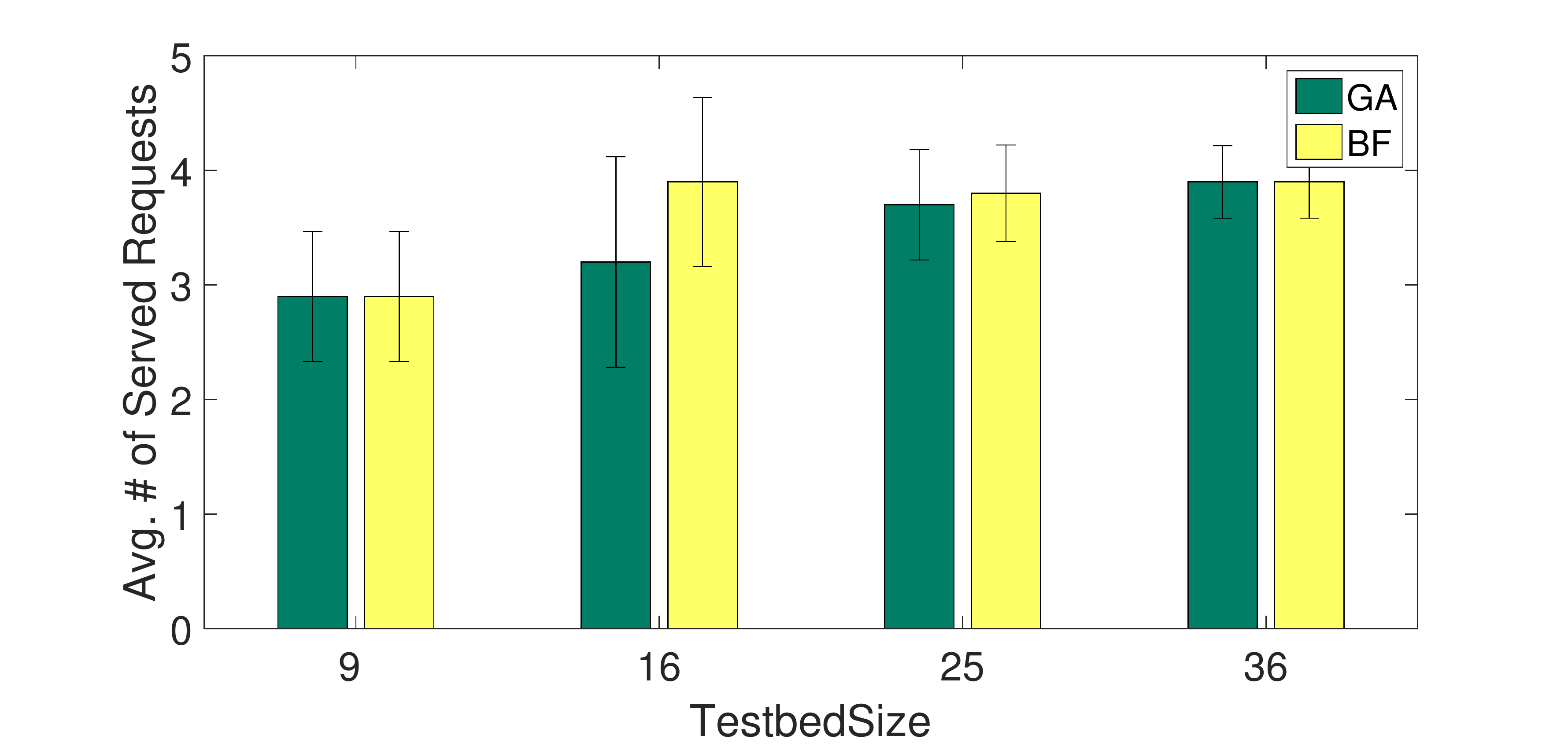}
		\caption{Virtual requests enabled, 15 Mappings.}
		\vspace{-1mm}
	\end{subfigure}
	\caption{\textit{1\textsuperscript{st} scenario I: Average number of served requests for grid topologies.} Mapping limits of 10 and 15, results, for requests of both virtualization enabled and of only physical nodes.} 
	\label{Served_grid}
	\vspace{-5mm}
\end{figure*}

\begin{figure*}[t]
	\centering
	\begin{subfigure}[t]{0.45\textwidth}
		\centering
		\includegraphics[width=\textwidth]{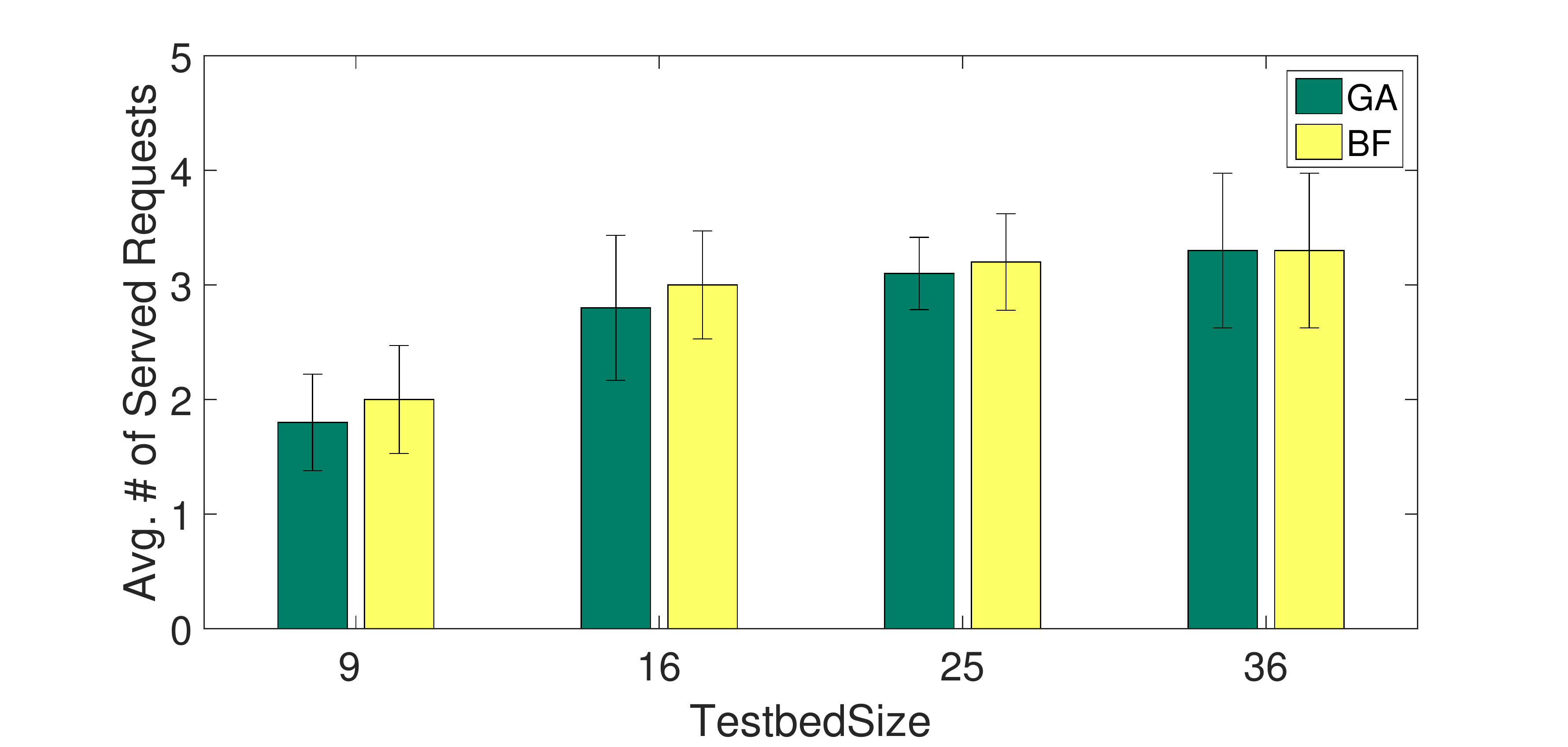}
		\caption{Physical requests, 10 Mappings.}
	\end{subfigure}
	\begin{subfigure}[t]{0.45\textwidth}
		\centering
		\includegraphics[width=\textwidth]{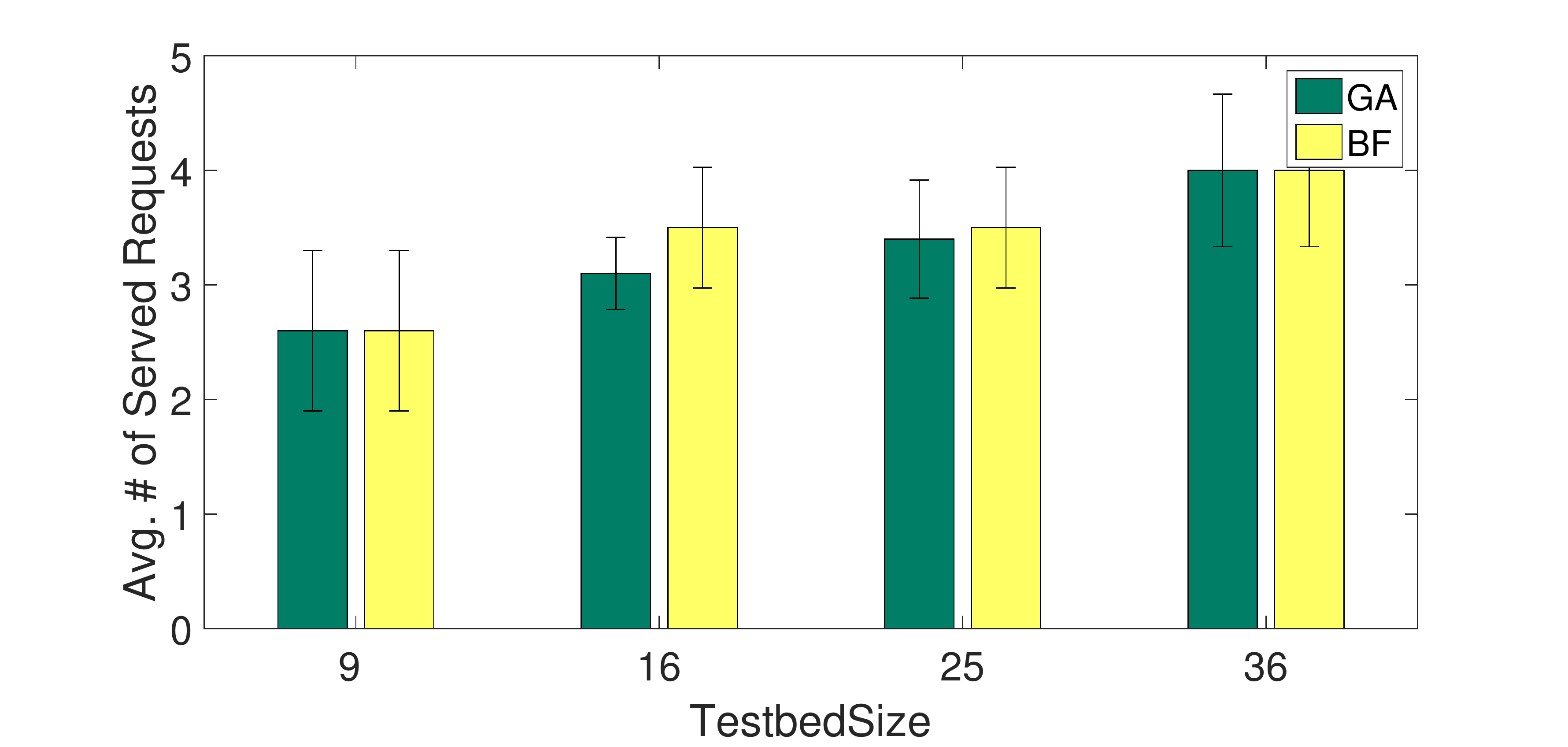}
		\caption{Virtual requests enabled, 10 Mappings.}
	\end{subfigure}	
	\begin{subfigure}[t]{0.45\textwidth}
		\centering
		\includegraphics[width=\textwidth]{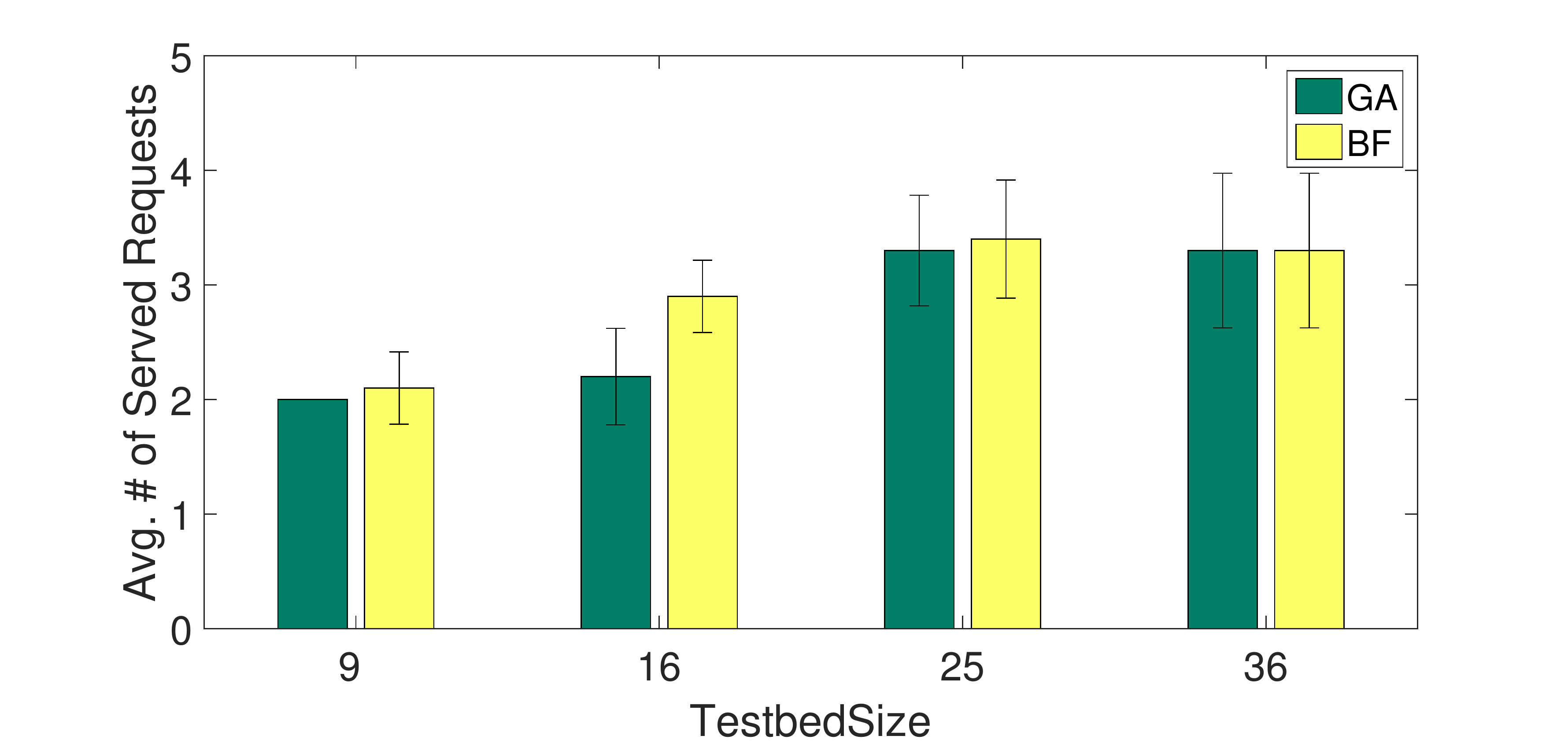}
		\caption{Physical requests, 15 Mappings.}
		\label{fig3:c}
		\vspace{-1mm}
	\end{subfigure}
	\begin{subfigure}[t]{0.45\textwidth}
		\centering
		\includegraphics[width=\textwidth]{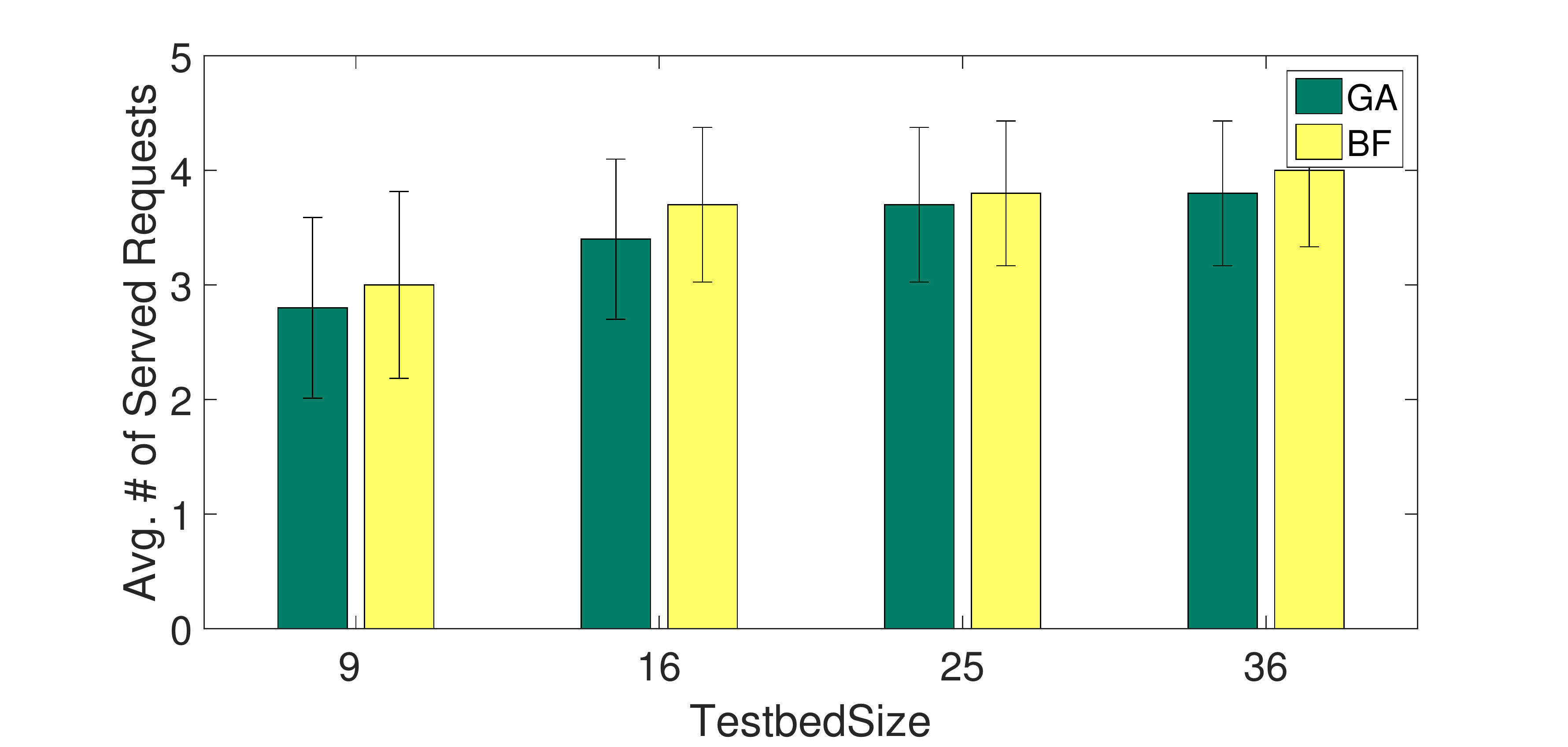}
		\caption{Virtual requests enabled, 15 Mappings.}
		\vspace{-1mm}
	\end{subfigure}
	\caption{\textit{1\textsuperscript{st} scenario II: Average number of served requests for random topologies.} Mapping limits of 10 and 15, results, for requests of both virtualization enabled and of only physical nodes.} 
	\label{Served_Rand}
	\vspace{-3mm}
\end{figure*}

\begin{figure}[t]
	\vspace{-3mm}
	\centering
	\includegraphics[width=3in]{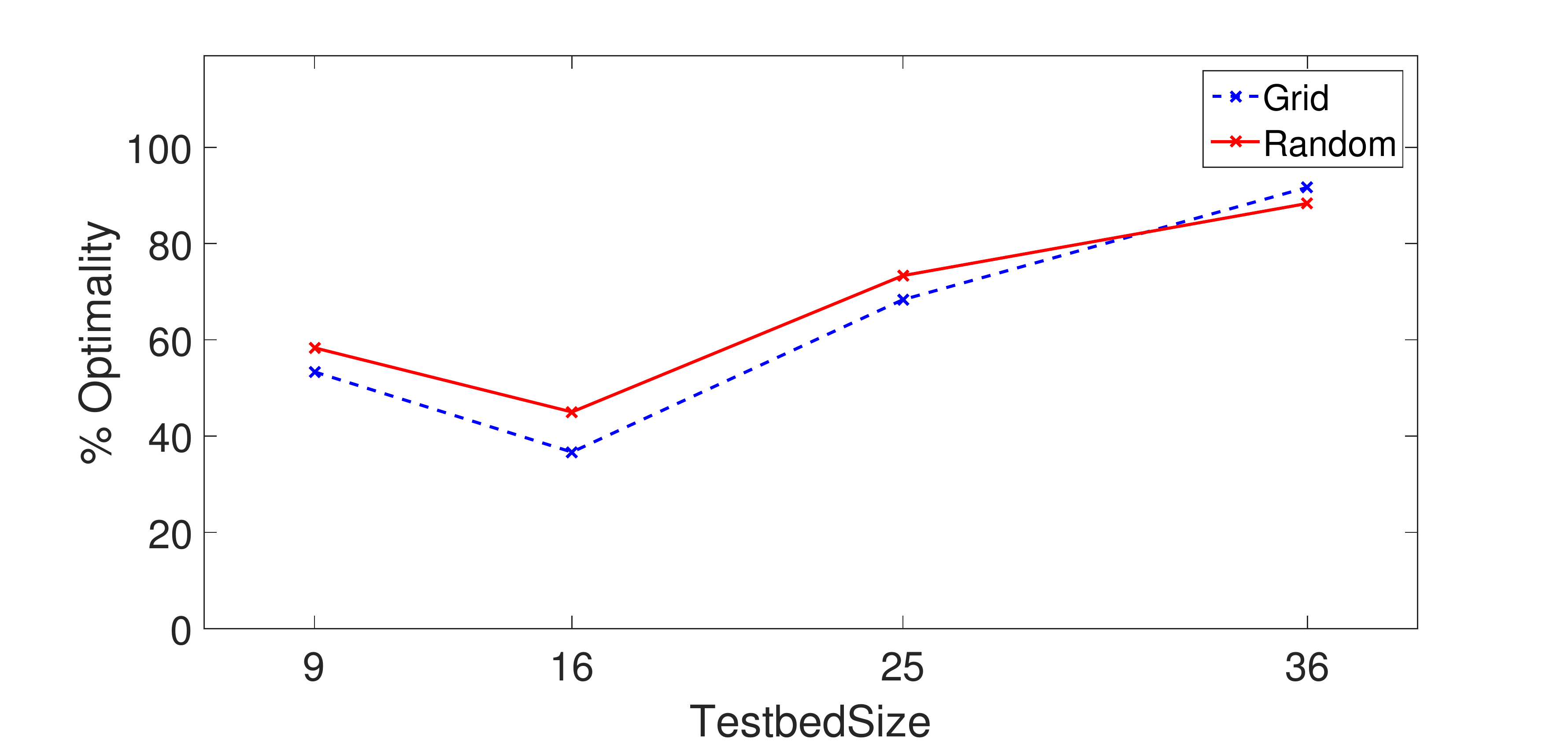}
	\vspace{-1mm}
	\caption{Percentage of optimal solutions found by GA-mapper on grid and random topologies.}
	\label{Opt_mean}
	\vspace{-4mm}
\end{figure}

\vspace{-1mm}
\subsection{Performance results}
\vspace{-1mm}
\figurename \ref{Served_grid} shows the average number of requests to be served for different mapping limits of 10 and 15, for both $Type = 1$ (virtualization-enabled) and $Type = 0$ (only physical) on the grid testbeds. In addition, \figurename \ref{Served_Rand} presents the results for randomly generated topologies of the same-sizes testbeds. We notice three bars on \figurename \ref{fig2:a}, \ref{fig2:c} and \figurename \ref{fig3:c}, with no error bars, where the results over ten runs exhibits no variations. As seen from the sub-figures, our mapper (GA) serves almost exactly the same number of requests compared to the brute force (BF) solution. In total, for about \textbf{82.96\%} of runs, the GA-based mapper serves the same number of requests as the BF. Besides, it is noticeable that the average number of served requests is not strictly increasing with the increase of testbed sizes, which can be attributed to the grid topology limited network connectivity as well as the high average number of requested channels per user. Furthermore, the number of successfully allocated requests increase with the mapping limits, which is intuitive. The more possible different mappings that are fed to the system, the more probable it could find a better solution and allocate more users concurrently. Moreover, the average number of allocated requests on the physical case is more than one, which demonstrates the sharing benefits of the frequency slicing technique used.\\
\begin{figure}[t]
	\centering
	\includegraphics[width=1.75in]{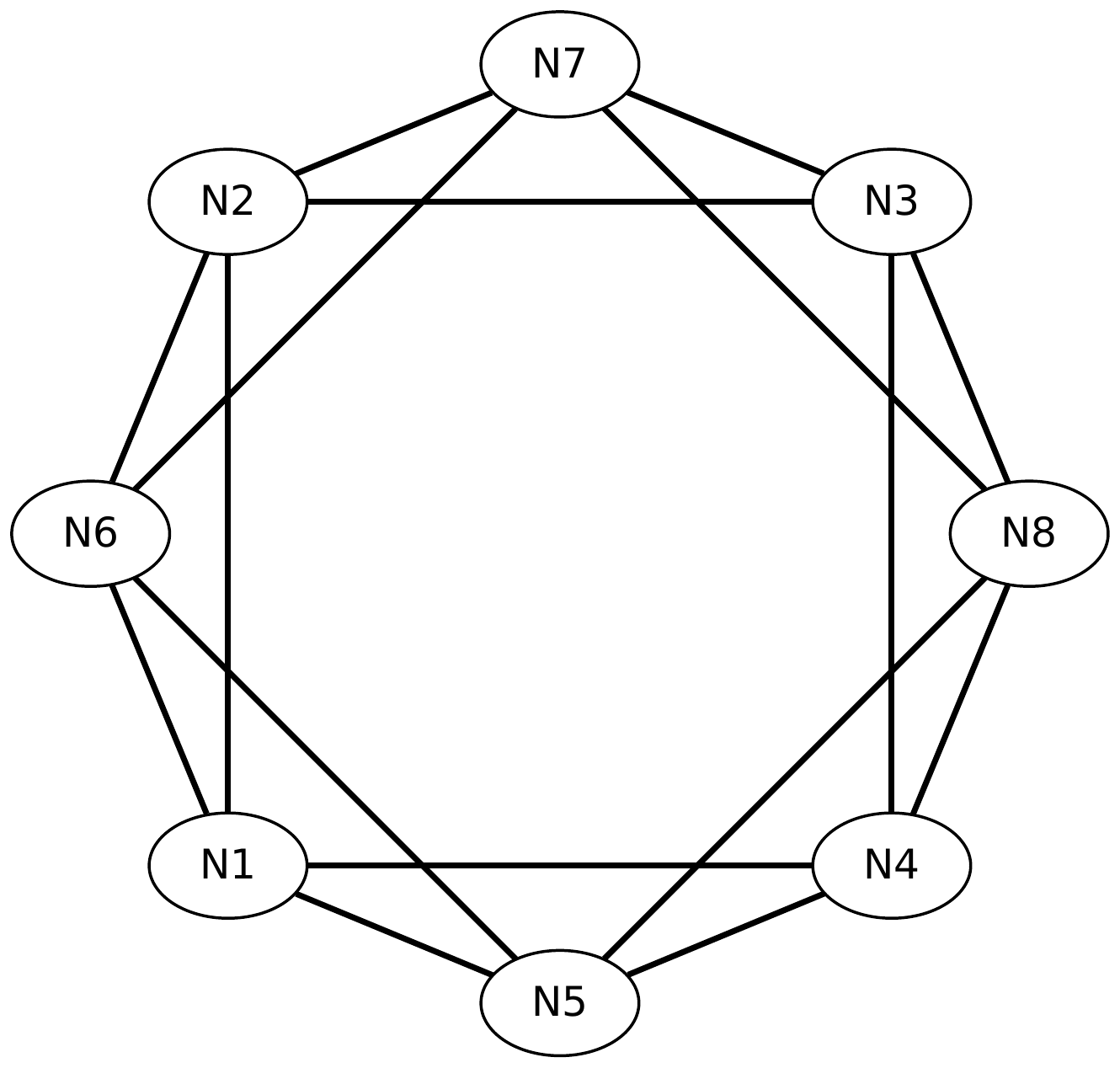}
	\vspace{-1mm}
	\caption{The topology of the adopted 8-nodes wireless testbed.}
	\label{topology}
	\vspace{-5mm}
\end{figure}
To demonstrate the effect of the testbed physical topology on the results of the average number of allocated requests, the same procedure is performed on carefully planned topology similar to the one used by the CRC testbed \cite{CRCtestbed}. The nodes graph of this 8-nodes network is shown on \figurename \ref{topology}. The results obtained on this topology, with five requests with the same parameters, and mapping limits of 10 and 15 are displayed on \figurename \ref{8nodes}. As observed from the figure, applying our mapper on this topology results in increase in number of served requests for a smaller sized testbed. This highlights the fact that the grid topology is not the best way to organize testbed nodes.\\
\figurename \ref{Opt_mean} shows the optimality metric results for both grid and random topologies testbeds with variable number of nodes in the testbed. The figure indicates a good steady performance with increased testbed sizes, because of the careful design of chromosomes embedding we used, with an average of \textbf{65\%} over all runs.\\
We manifest the slicing revenue resulted from resources slicing and our mapping algorithm on the testbed utilization on \figurename \ref{Revenues_grid} and \figurename \ref{Revenues_Rand},  using all mappings of the first stage of our approach. The figures show the effect of varying the incoming number of requests on the revenue, and the impact of varying the testbed size as well for both grid and random topologies.  
Our adopted frequency-based slicing scheme considered by our mapper offers up to 490\% increase in the number of concurrent users relative to the simple allocation scheme where no slicing supported on grid topologies, and up to 500\% increase on random topologies.\\
Finally and as expected, our algorithm solves this optimization problem pretty fast relative to the exhaustive (BF) search algorithm. As depicted in \figurename \ref{time_fig}, the exhaustive search execution time increases exponentially with the increase in search space (mappings limit) while our algorithm scales almost linearly. 

\begin{figure}[t]
	\centering
	\vspace{-4mm}
	\includegraphics[width=3in]{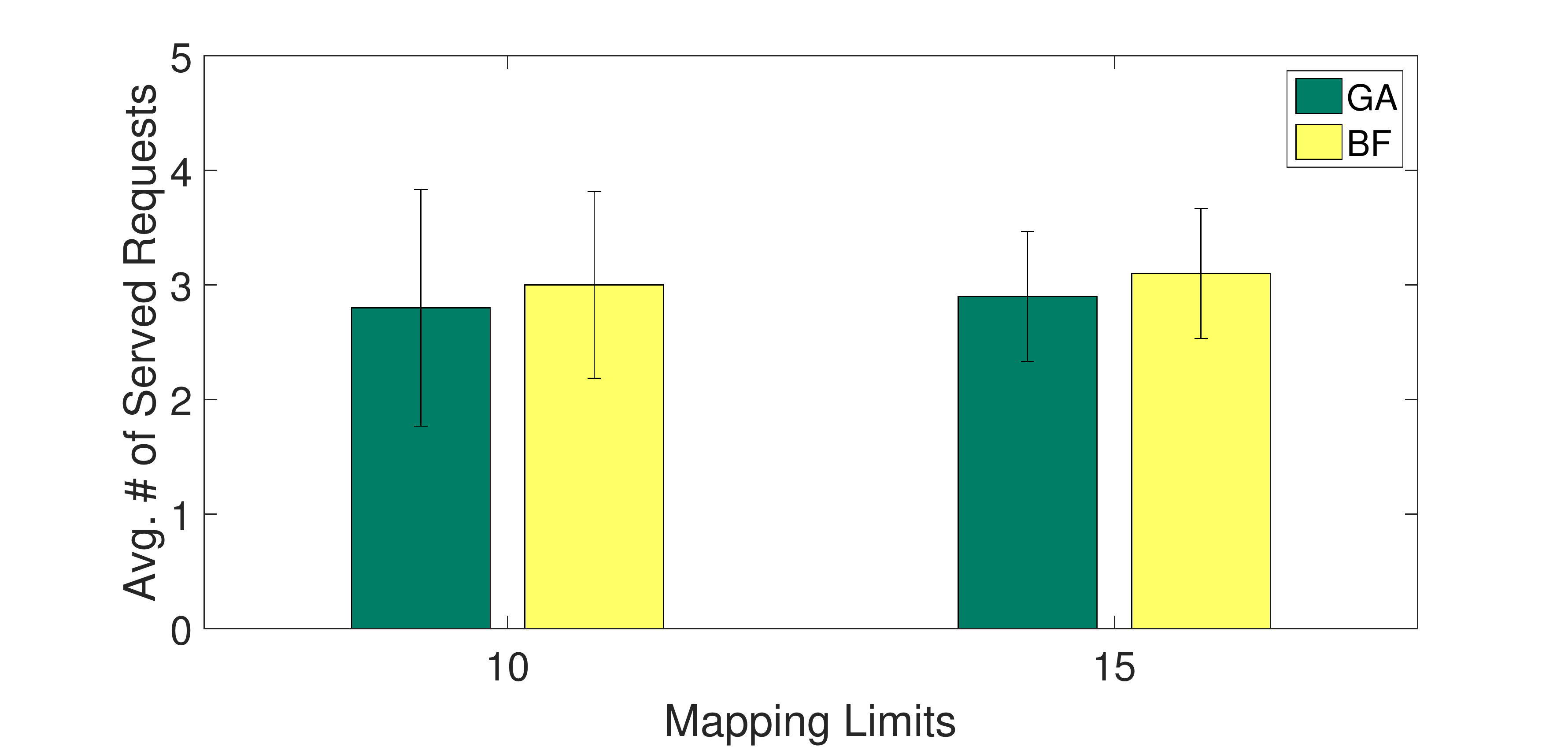}
	\vspace{-1mm}
	\caption{Average number of served requests for mapping limits of 10 and 15 for the 8-node testbed, virtualization enabled.}
	\vspace{-3mm}
	\label{8nodes}
\end{figure}


\begin{figure}[t]
	\centering
	\includegraphics[width=3in]{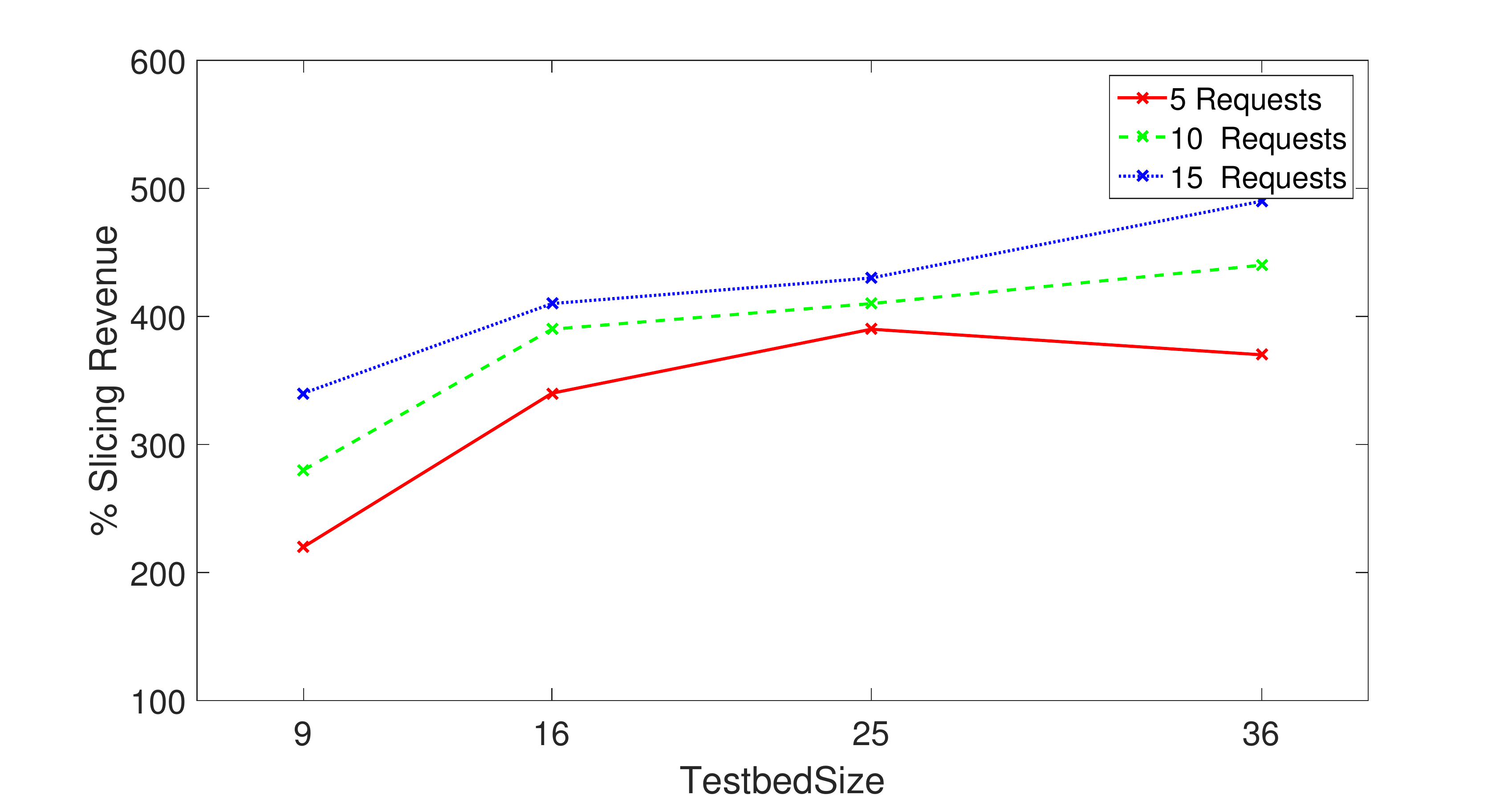}
	\vspace{-1mm}
	\caption{Slicing Revenue for different number of requests and different grid testbed sizes.}
	\vspace{-3mm}
	\label{Revenues_grid}
\end{figure}

\begin{figure}[t]
	\centering
	\includegraphics[width=3in]{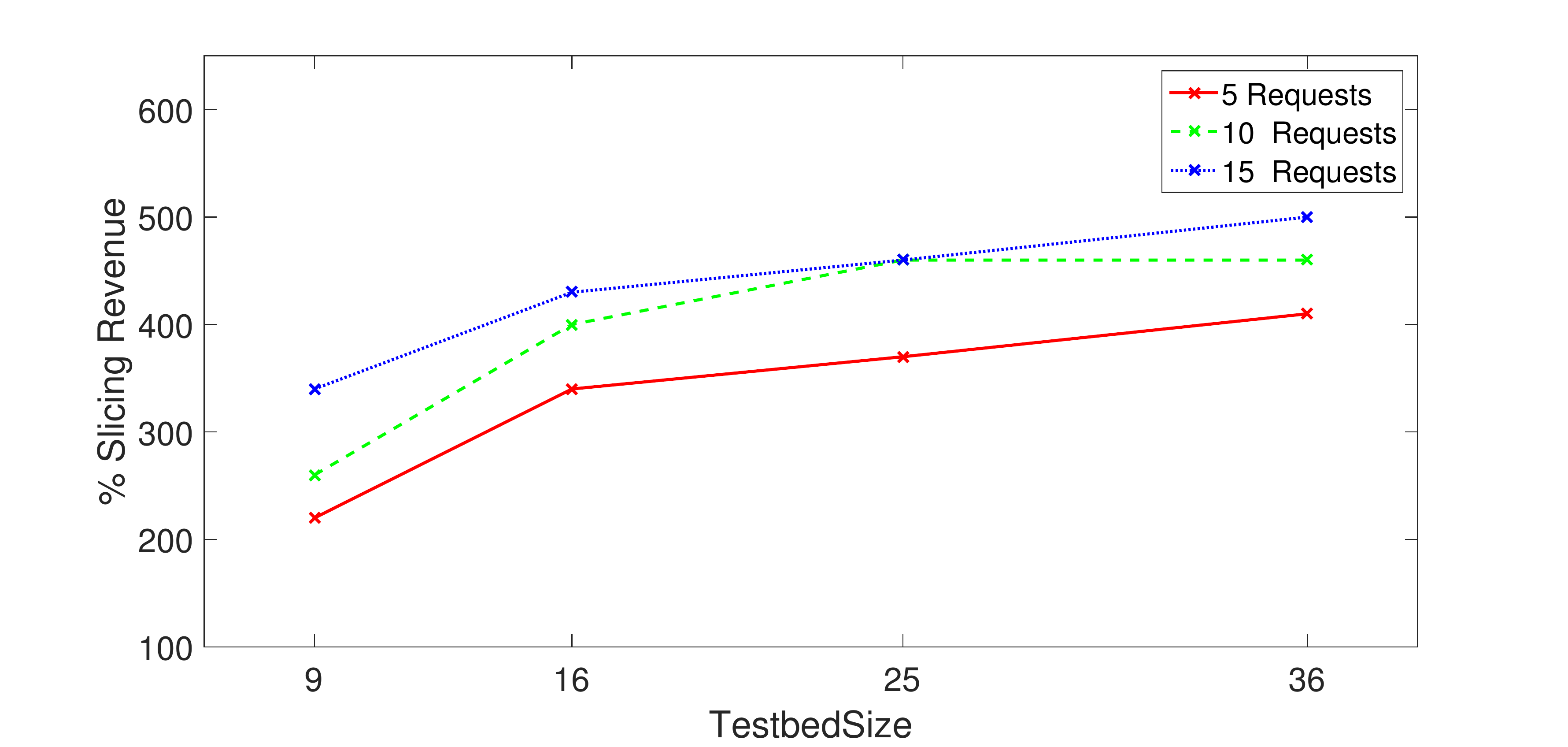}
	\vspace{-1mm}
	\caption{Slicing Revenue for different number of requests and different testbed sizes of random topologies.}
	\vspace{-3mm}
	\label{Revenues_Rand}
\end{figure}


\begin{figure}[t]
	\centering
	\includegraphics[width=3in]{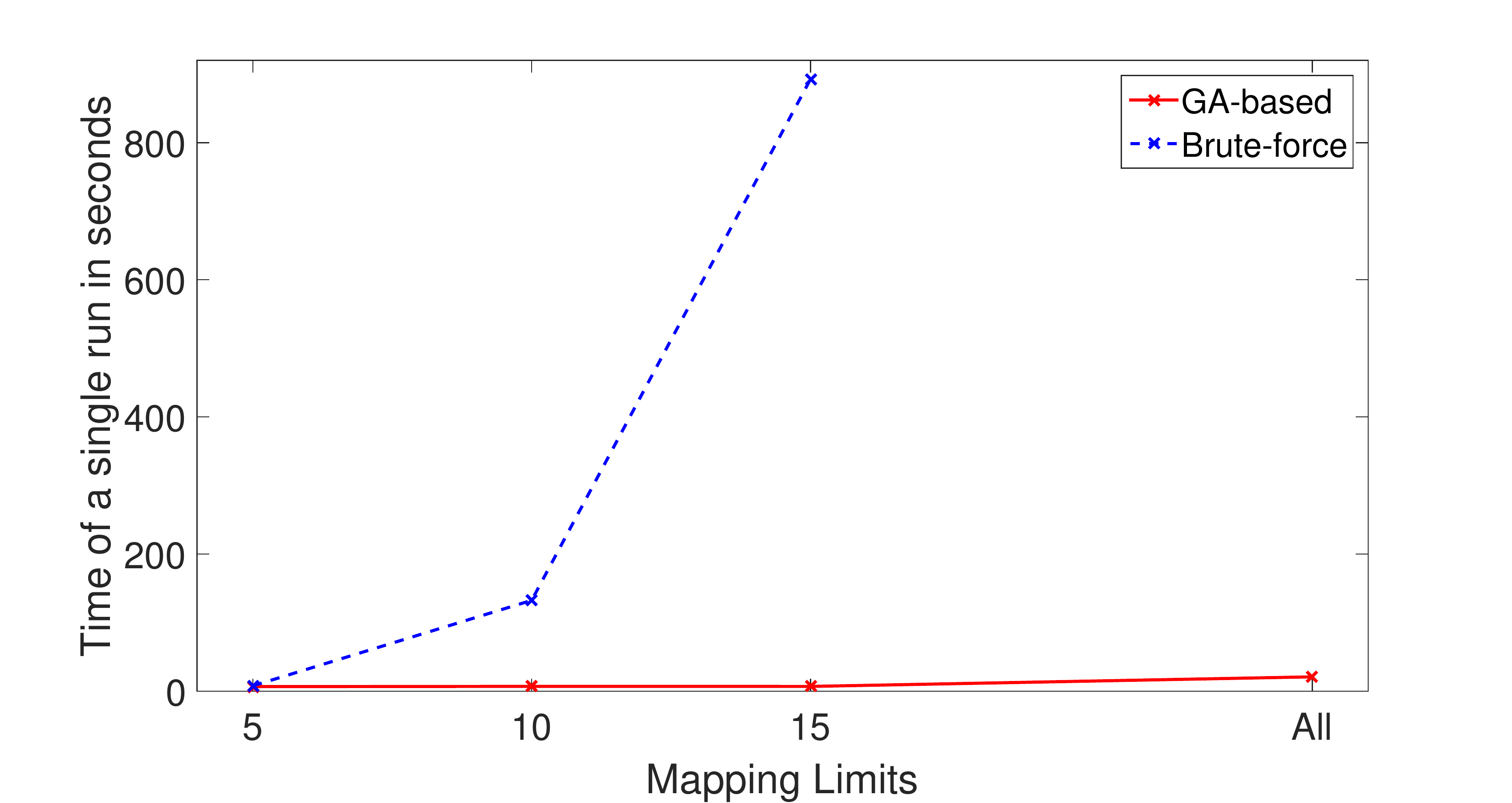}
	\vspace{-1mm}
	\caption{Elapsed execution time of brute-force versus GA-based mapper for different mapping limits.}
	\vspace{-6mm}
	\label{time_fig}
\end{figure}

\vspace{-0.5mm}
\section{Conclusion and Future Work}
\label{sec:conc}
\vspace{-1mm}
This paper presents our implementation of a mapping algorithm for reservation requests on wireless testbeds, based on induced sub-graph isomorphism and genetic algorithm. We describe the concept and illustrate its performance using simulation for different scenarios.
In almost all of the experiments performed, our genetic algorithm-based mapper \textit{succeeded to allocate the same number of requests allocated by the best solution found by brute force search in \textbf{82.96\%} cases}. We manifested that a careful choice of nodes locations of the testbed can enhance the number of users that can be concurrently served. Moreover, our proposed technique achieves up to 500\% revenue of testbed users by the suggested slicing mechanism compared to no-slicing strategy.\\
We are working on integrating a 24 hours scheduler into the current implementation of mapping algorithm. The fitness function will be re-designed to include scheduling details, therefore, performing both mapping and providing the daily schedule of requests for each time-slot.   



\vspace{-1mm}
\bibliographystyle{IEEEtran}
\vspace{-1mm}
\bibliography{IEEEabrv,Final_2}

%



\end{document}